\documentclass[twocolumn,showpacs,preprintnumbers,floatfix,
               amsmath,amssymb]{revtex4}
\usepackage{graphics}                           
\usepackage{epsfig}                             
\begin{document}

\title{Spin-orbit splitting in low-{\it j} neutron orbits \\
       and proton densities in the nuclear interior}

\author{B.G. Todd, J. Piekarewicz and P.D. Cottle}

\affiliation{Department of Physics, Florida State University,
Tallahassee, Florida 32306}


\date{\today}

\begin{abstract}
On the basis of relativistic mean field calculations, we predict that
the spin-orbit splitting of $p_{3/2}$ and $p_{1/2}$ neutron orbits
depends sensitively on the magnitude of the proton density near the
center of the nucleus, and in particular on the occupation of
$s_{1/2}$ proton orbits. We focus on two exotic nuclei, $^{46}$Ar and
$^{206}$Hg, in which the presence of a pair of $s_{1/2}$ proton holes
is predicted to cause the splitting between the $p_{3/2}$ and
$p_{1/2}$ neutron orbits near the Fermi surface to be much smaller
than in the nearby doubly-magic nuclei $^{48}$Ca and $^{208}$Pb.
We note that these two exotic nuclei depart from the long-standing
paradigm of a central potential proportional to the ground state 
baryon density and a spin-orbit potential proportional to the 
derivative of the central potential.
\end{abstract}

\maketitle

One of the primary motivations for the study of exotic nuclei is to
search for novel shell structure effects. A large amount of attention
has been paid to the possibility that the spin-orbit force on high-$j$
neutron orbits weakens in nuclei near the neutron drip line
\cite{We96,Te97,Re97,La98,De99,Du99,Re99,La99,Ca00,Pe00,De01,Ot01,Mi02,Ro02}.
The neutron magic numbers for stable nuclei rely on the effect of the
strong spin-orbit force on high-$j$ orbits, so the weakening of this
force has the potential to change the neutron magic numbers in
neutron-rich nuclei. The possibility of the narrowing or collapse of
the $N=28$ major shell closure in neutron-rich nuclei near $^{42}$Si
has attracted considerable attention because these isotopes are
becoming accessible to experiments~\cite{Gl97,Sa00,So02,No02}. The
two most important reasons generally given for the decline of the
spin-orbit force on high-$j$ neutron orbits near the neutron drip line
are the large neutron surface diffuseness and the influence of the
continuum in these nuclei \cite{Na01,Ca00b}.

In the present communication, we predict a novel shell structure
effect having to do with spin-orbit splitting in {\it low-j} neutron
orbits --- namely, {\it p} orbits.  The dramatic decrease in the
spin-orbit splitting described here is {\it not} caused by the neutron
density near the nuclear surface, but rather by the {\it proton density
in the nuclear interior}.  The two nuclei for which we make specific
predictions, $^{46}$Ar and $^{206}$Hg, are exotic but within two
protons of the valley of stability.  We make these predictions using
the relativistic mean field theory, which has also been used to study
the spin-orbit splitting of high-{\it j} orbits in exotic nuclei
\cite{La98,La99,Ca00,De01,Mi02}.

The relativistic mean field calculation reported here is identical to
the calculation used in Ref.~\cite{To03} to predict the properties of
neutron-rich nuclei over a wide mass range.  The model used in
Ref.~\cite{To03} is based on a Lagrangian developed in
Refs.~\cite{Ho01a,Ho01b} that includes novel nonlinear
couplings between the isoscalar and isovector mesons. These new
terms, which supplement the phenomenologically successful Lagrangians
of Refs.~\cite{La97,La99b,Mu96}, modify the density dependence of the
symmetry energy without changing ground state properties that are well
established experimentally. Modifications to the poorly known density
dependence of the symmetry energy induces interesting correlations 
between the neutron skin of heavy nuclei and a variety of neutron-star 
properties~\cite{Ho01a,Ho01b,Ho02,Ca03}.

\begin{figure}[ht]
\begin{center}
\includegraphics[width=3.0in,angle=0,clip=false]{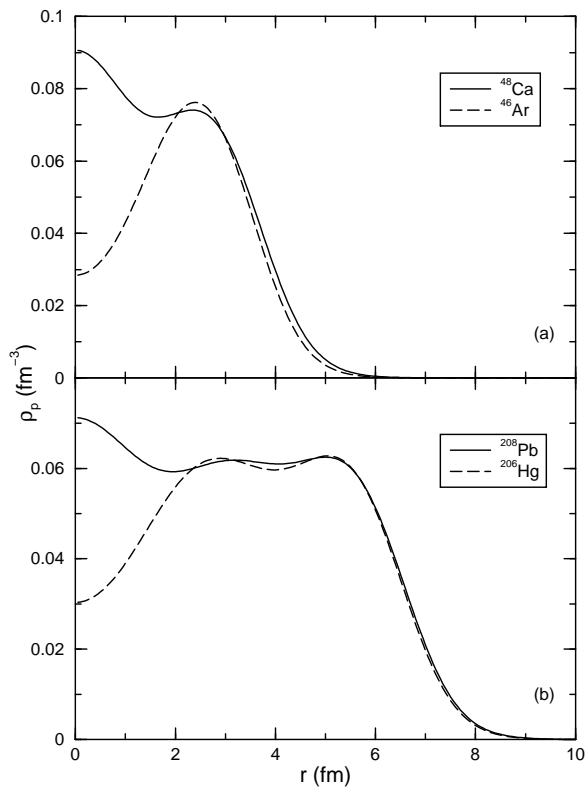}
\caption{Proton (point) densities for (a) $^{46}$Ar and $^{48}$Ca 
and for (b) $^{206}$Hg and $^{208}$Pb 
computed using the relativistic parametrization of Ref.~\cite{La97}. 
The development of a ``proton hole'' in the interior of the nucleus
is readily observed.}
\label{Fig1}
\end{center}
\end{figure}

\begin{figure}[ht]
\begin{center}
\includegraphics[width=3.0in,angle=0,clip=true]{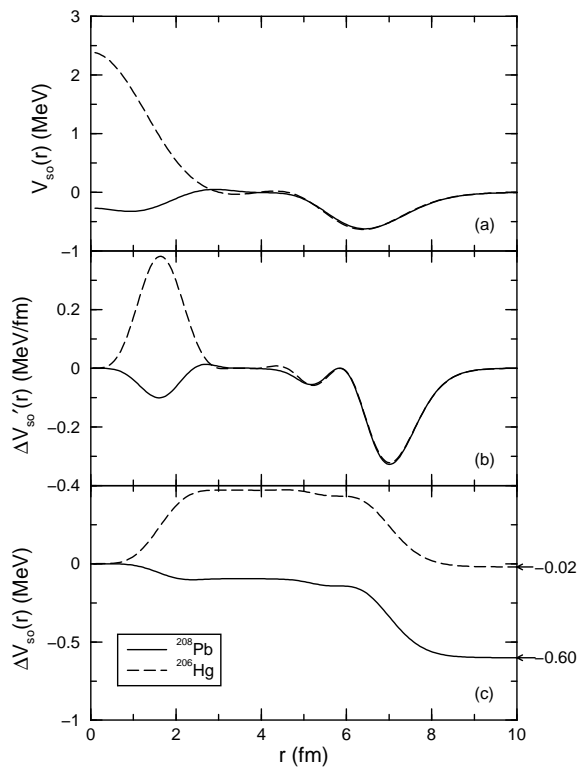}
\caption{(a) The Schr\"odinger-equivalent spin-orbit potential
         for $^{208}$Pb (solid line) and $^{206}$Hg (dashed
	 line). Panels (b) and (c) display the effect of
	 folding the spin-orbit potential with the 
	 Schr\"odinger-equivalent $p$-orbitals, as defined in
	 Eq.~(\ref{Vspinorbit}). The arrows point to a first-order 
	 estimate of the spin-orbit splitting.}
\label{Fig2}
\end{center}
\end{figure}

In both doubly-magic nuclei $^{48}$Ca and $^{208}$Pb, the highest
lying proton orbits below the Fermi surface (or the lowest energy
proton hole states in $^{47}$K and $^{207}$Tl) are $s_{1/2}$ orbits.
The effect of removing a pair of $s_{1/2}$ protons from $^{48}$Ca and
$^{208}$Pb is illustrated in Fig.~\ref{Fig1}, which compares the
proton densities of $^{46}$Ar and $^{48}$Ca (upper panel), and the
proton densities of $^{206}$Hg and $^{208}$Pb (lower panel).  It is
important to note that the predicted root-mean-square charge radii for
$^{48}$Ca and $^{208}$Pb are in excellent agreement with
experiment~\cite{Vr87}.  As the $s_{1/2}$ wavefunctions are strongly
peaked in the center of the nucleus, the removal of these protons from
$^{48}$Ca and $^{208}$Pb results in sharply reduced proton densities
in the centers of $^{46}$Ar and $^{206}$Hg.  This, in turn, causes a
sharp increase in the magnitude of the spin-orbit interaction in the
nuclear interior.  Figure~\ref{Fig2}(a) illustrates this effect in
$^{208}$Pb and $^{206}$Hg; inside of 2 fm, $V_{\rm so}$ is much
stronger --- and of the opposite sign --- in the $s_{1/2}^{-2}$
nucleus $^{206}$Hg than in the doubly-magic $^{208}$Pb core.
This unconventional behavior of the spin-orbit potential is intimately 
related to the Lorentz structure of the Dirac mean fields. While the 
depletion of $s_{1/2}$ proton strength manifests itself in both the 
vector and scalar densities, the ``proton-hole'' disappears from the 
central potential as a result of the sensitive cancellation between the 
attractive scalar and the repulsive vector potentials~\cite{Se86}. 
In contrast, (the derivatives of) the scalar and vector potentials 
add constructively in the spin-orbit potential and the development 
of a nontrivial spin-orbit structure in the interior of the nucleus 
ensues.

Figures~\ref{Fig2}(b)-(c) illustrate the effect of folding the 
spin-orbit potential in $^{208}$Pb and $^{206}$Hg with the 
$3p_{1/2}$ and $3p_{3/2}$ neutron wavefunctions. 
That is, we display
\begin{equation}
  \Delta V_{\rm so}(r)\equiv \int_{0}^{r}dr' V_{\rm so}(r')
  \left[2u_{p_{3/2}}^{2}(r')+u_{p_{1/2}}^{2}(r')\right]\;.
 \label{Vspinorbit}
\end{equation}
Expressions for the Schr\"odinger-equivalent spin-orbit 
potential and wavefunctions [$u(r)$] may be found in 
Ref.~\cite{Pi93}. For the purposes of this study, normalized 
wavefunctions have been used.  Note that the above quantity, while 
not exact, provides an accurate (first-order) estimate 
of the $p_{3/2}$--$p_{1/2}$ spin-orbit splitting 
$\Delta V_{\rm so}\!\equiv\!
\Delta V_{\rm so}(r\!\rightarrow\!\infty)$.

\begin{figure}[ht]
\begin{center}
\includegraphics[width=3.0in,angle=0,clip=true]{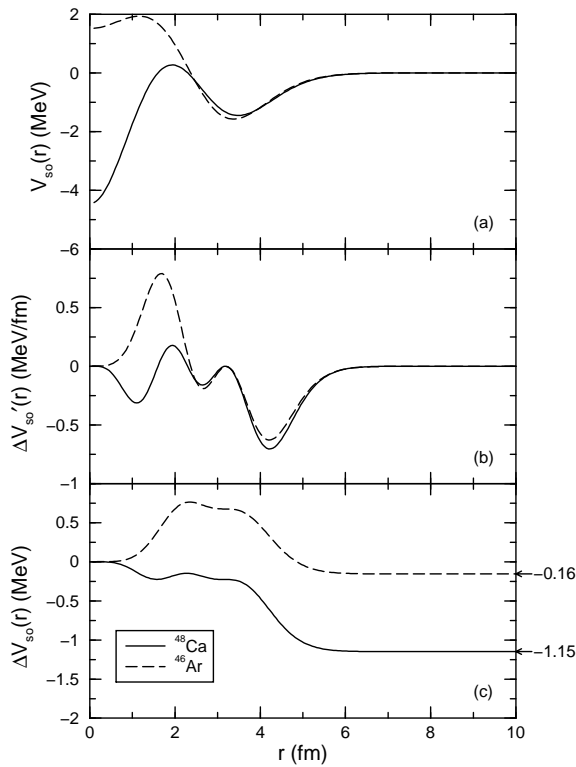}
\caption{(a) The Schr\"odinger-equivalent spin-orbit potential
         for $^{48}$Ca (solid line) and $^{46}$Ar (dashed
	 line). Panels (b) and (c) display the effect of
	 folding the spin-orbit potential with the 
	 Schr\"odinger-equivalent $p$-orbitals, as defined in
	 Eq.~(\ref{Vspinorbit}). The arrows point to a first-order 
	 estimate of the spin-orbit splitting.}
\label{Fig3}
\end{center}
\end{figure}

\begin{figure}[ht]
\begin{center}
\includegraphics[width=3.0in,angle=0,clip=true]{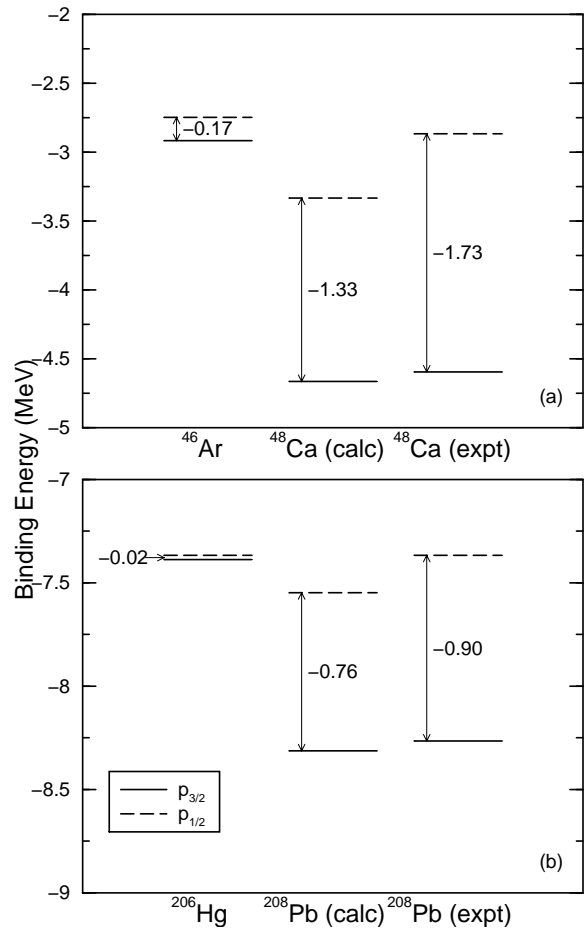}
\caption{Comparison between the experimental and calculated
         $p_{3/2}$-$p_{1/2}$ spin-orbit splitting for the 
         doubly-magic nuclei $^{48}$Ca and $^{208}$Pb. Also
	 shown is the predicted collapse of the spin-orbit
	 splitting in the two exotic nuclei $^{46}$Ar and
	 $^{206}$Hg.}
\label{Fig4}
\end{center}
\end{figure}

The combined effect of a strong increase in $V_{\rm so}$ in the
interior of $^{206}$Hg together with neutron wavefunctions that 
are much larger at small radii than larger $j$ orbits, yields a 
big effect on the integrated spin-orbit energy and, therefore, 
on the spin-orbit splitting for the $p$--neutrons. Indeed, this 
effect leads to the collapse of the $p_{3/2}$--$p_{1/2}$
spin-orbit splitting: from $-0.60$ MeV in $^{208}$Pb to $-0.02$ 
MeV in $^{206}$Hg. A similar effect occurs in $^{46}$Ar relative 
to $^{48}$Ca: the spin-orbit splitting is reduced by almost an 
order of magnitude, as shown in Fig.~\ref{Fig3}. Figure~\ref{Fig4} 
displays the exact values. We conclude that in the two exotic 
nuclei $^{46}$Ar and $^{206}$Hg, the spin-orbit interaction ceases 
to be a surface-dominated phenomenon.

Figure~\ref{Fig4} summarizes the calculations and compares the
experimental and calculated binding energies for $p_{3/2,1/2}$ neutron
orbits in $^{48}$Ca and $^{208}$Pb. It should be emphasized that while
models with nonlinear couplings between the isoscalar and isovector
mesons change the energy of the individual $p_{3/2}$ and $p_{1/2}$
orbitals slightly, the prediction for their spin-orbit splitting is
largely model independent. Thus, we limit ourselves to the original
NL3 set of Ref.~\cite{La97}. The experimental binding energies for
$^{48}$Ca are taken from the $^{48}$Ca($d$,$p$) measurement of Uozumi
{\it et al.}~\cite{Uo94} and the mass compilation of Audi and
collaborators~\cite{Au97}. The $^{208}$Pb($p$,$d$) data used to
extract the binding energies for $^{208}$Pb are taken from the
compilation of Martin~\cite{Ma93}; the mass data are taken from
Ref.~\cite{Au97}.

For stable nuclei, the standard experimental technique for mapping
single neutron strength in a nucleus is to use a stripping reaction
such as ($d$,$p$). To differentiate between spin-orbit partners (such
as $p_{3/2}$ and $p_{1/2}$) a polarized deuteron beam would be used
(as in \cite{Uo94}). For the exotic nucleus $^{46}$Ar, such a
measurement would be performed in inverse kinematics with a $^{46}$Ar
beam and polarized deuteron target. The measurement would further be
complicated by the likelihood that the $p_{3/2}$ and $p_{1/2}$
strengths would be somewhat fragmented, as they are in $^{51}$Ti,
$^{53}$Cr and $^{55}$Fe~\cite{Ko72}. In $^{206}$Hg, the $p_{3/2,1/2}$
orbits would be observed as holes, requiring the use of the pickup
reaction ($p$,$d$) in inverse kinematics, once again with a polarized
target to differentiate between spin-orbit partners. For example, the
normal kinematics experiment $^{208}$Pb($p$,$d$) with a polarized beam
is reported in \cite{Ma97}.

In summary, we have used relativistic mean field calculations to
predict that the spin-orbit splitting of $p_{3/2}$ and $p_{1/2}$
neutron orbits depends sensitively on the magnitude of the proton
density near the center of the nucleus, and in particular on the
occupation of $s_{1/2}$ proton orbits.  The quenching (or collapse) of
the spin-orbit splitting in high-$j$ neutron orbits has been
advertised as the hallmark for novel nuclear-structure effects in
neutron-rich nuclei. This collapse is associated with the development
of a diffuse neutron-rich surface. In this communication we have
proposed a new mechanism for the collapse of the spin-orbit splitting
--- but among low-$j$ neutron orbits. This mechanism is based, not on
a rearrangement of the neutron density at the surface of the nucleus,
but rather, on a depletion of the proton density in the nuclear
interior.  Two exotic nuclei, $^{46}$Ar and $^{206}$Hg, may be
accessible for the study of this effect. In these nuclei we predict
that the presence of a pair of $s_{1/2}$ proton holes causes the
splitting between the $p_{3/2}$ and $p_{1/2}$ neutron orbits near the
Fermi surface to be much smaller than in the nearby doubly-magic
nuclei $^{48}$Ca and $^{208}$Pb.
Thus these two exotic nuclei, only two protons away from being doubly
magic, deviate from a long-standing paradigm that has been applied
with enormous success in both structure and reaction calculations,
namely, that of a central potential proportional to the ground state
baryon density and a spin-orbit potential proportional to the
derivative of the central potential. Finally, while the effects
proposed here are likely to be modified by correlations that go beyond
the mean-field approximation, we trust that most of its novel
qualitative features will hold true.

This work was supported in part by the U.S. Department of Energy 
through grant DE-FG05-92ER40750, the National Science Foundation 
through grant PHY-0139950, and the State of Florida.

\end{document}